\documentclass[letterpaper, 10 pt, conference]{ieeeconf}
\IEEEoverridecommandlockouts
\usepackage{cite}
\usepackage[letterpaper, top=57pt, left=52pt, right=52pt, bottom=57pt]{geometry}
\usepackage{amsmath,amssymb,amsfonts}
\usepackage{algorithmic}
\usepackage{graphicx}
\usepackage{textcomp}
\usepackage{xcolor}
\usepackage{dsfont}
\def\BibTeX{{\rm B\kern-.05em{\sc i\kern-.025em b}\kern-.08em
    T\kern-.1667em\lower.7ex\hbox{E}\kern-.125emX}}

\newcommand{\n}{\mathbb{N}}

\newcommand{\bP}{\mathbb{P}}
\newcommand{\BB}{\mathcal{B}}
\newcommand{\CC}{\mathcal{C}}
\newcommand{\DD}{\mathcal{D}}

\newcommand{\UU}{\mathcal{U}}

\newcommand{\XX}{\mathcal{X}}

\newcommand{\real}{\mathbb{R}}

\newcommand{\EE}{\mathbb{E}}

\newcommand\setb[1]{\{#1\}}
\newcommand{\setcomp}[1]{\overline{#1}} 

\newcommand{\ra}{\rightarrow}
\newcommand{\ie}{\text{i.e. }}
\newcommand{\st}{\text{s.t. }}

\newcommand*{\tran}{^{\mkern-1.5mu\mathsf{T}}}

\newtheorem{definition}{Definition}

\newtheorem{assumption}{Assumption}
\newtheorem{remark}{Remark}
\newtheorem{proposition}{Proposition}

\newtheorem{problem}{Problem}
\newtheorem{theorem}{Theorem}

\begin{document}

\title{Safety Verification of Stochastic Systems under Signal Temporal Logic Specifications
}

\author{Liqian Ma, Zishun Liu, Hongzhe Yu, and Yongxin Chen
\thanks{The authors are with Georgia Institute of Technology, Atlanta, GA 30332 
        {\tt\small \{mlq\}\{zliu910\}\{hyu419\}\{yongchen\}@gatech.edu}}%
}


\maketitle

\begin{abstract}
We study the verification problem of stochastic systems under signal temporal logic (STL) specifications. We propose a novel approach that enables the verification of the probabilistic satisfaction of STL specifications for nonlinear systems subject to both bounded deterministic disturbances and stochastic disturbances. Our method, referred to as the STL erosion strategy, reduces the probabilistic verification problem into a deterministic verification problem with a tighter STL specification. The degree of tightening is determined by leveraging recent results on bounding the deviation between the stochastic trajectory and the deterministic trajectory. Our approach can be seamlessly integrated with any existing deterministic STL verification algorithm. Numerical experiments are conducted to showcase the efficacy of our method.
\end{abstract}



\section{Introduction}
Safety is a critical consideration in various applications, including robots, autonomous vehicles, smart grids, and transportation control systems~\cite{wolf2017safety}. These safety-critical scenarios demand formal guarantees to ensure that systems operate as expected, as failures may result in severe consequences, such as harm to humans or significant financial costs. Safety verification refers to the task of determining whether a system satisfies a given safety specification over a specified period~\cite{guiochet2017safety, vicentini2019safety}. 
Conventional safe set specifications primarily focus on spatial requirements, ensuring that the system state never enters an unsafe region~\cite{prajna2004safety}. However, as the complexity of autonomous systems increases, many real-world tasks require specifications that are not only spatial but also temporal in nature. For instance, a mobile robot needs to pass Area A before entering Area B. In this paper, we focus on safety verification under Signal Temporal Logic (STL) specification, which uses both boolean and temporal logic operators to formulate constraints for continuous-valued systems~\cite{maler2004monitoring}. 

Real-world systems are subject to various types of uncertainty. It is essential for safety verification algorithms to account for disturbances. Many existing approaches model these uncertainties as bounded disturbances and employ worst-case analysis to guarantee the satisfaction of safety specifications. Examples of such approaches for safe set specification include Hamilton–Jacobi Reachability (HJ Reachability)~\cite{bansal2017hamilton}, reachability analysis, and barrier certificates~\cite{prajna2004safety}. For STL specifications, methods such as HJ Reachability~\cite{chen2018signal} and reachability analysis~\cite{roehm2016stl, lercher2024using, kochdumper2024fully} have been employed to formally verify STL satisfaction under bounded disturbance inputs.

In many practical situations, disturbances are better modeled as stochastic noise, which provides a more realistic representation, as in the case of sensor noise. When considering stochastic disturbances, the aforementioned deterministic methods are not applicable or tend to be overly conservative, as they focus on worst-case scenarios that rarely occur in practice. To better account for stochastic disturbances, we adopt a probabilistic setting, where the goal is to ensure the safety specification is satisfied with high probability, e.g., greater than 99.9\%. 
For safe set specifications, several methods have been proposed to verify stochastic systems, including martingale-based approaches~\cite{steinhardt2012finite, santoyo2021barrier}, risk estimation~\cite{frey2020collision}, and sampling-based methods~\cite{janson2017monte}. Our recent paper significantly reduces the conservativeness of the verification algorithms for safe set specifications~\cite{liu2024safety}.
For STL specifications, most existing approaches are limited to handling the probability constraint for a single trajectory satisfying the STL specification~\cite{sadigh2016safe, farahani2018shrinking, yang2023distributed, vlahakis2024probabilistic, kordabad2024control}. Very few studies have focused on STL verification under both bounded and stochastic disturbances. In \cite{salamati2021data}, a method is proposed to address this problem for linear systems under Gaussian noise. In this work, we focus on the problem of STL verification for nonlinear systems under both bounded and stochastic disturbances. 

In this work, we present a novel framework for verifying the probabilistic STL satisfaction of discrete-time nonlinear stochastic systems. To the best of our knowledge, this is the first approach capable of addressing this problem for nonlinear systems under both bounded and stochastic disturbances. Given a desired probability requirement, our method first erodes the superlevel set of the predicates in an STL formula to get a tighter STL formula. If the deterministic system is verified to satisfy the tighter STL formula, then the stochastic system is guaranteed to satisfy the original STL formula with the specified probability constraint. As a result, the stochastic verification problem is transformed into a deterministic one. The depth of erosion is determined by the sharp probabilistic bound proposed in our previous work~\cite{liu2024probabilistic}, which helps reduce the conservativeness of the verification result, especially when the probability tolerance is low and the time horizon is long. Our method does not rely on restrictive assumptions, such as linear system dynamics or affine predicates, which is common in previous work~\cite{vlahakis2024probabilistic}. This broader applicability makes our approach suitable for real-world applications.

\textit{Notations.}
Denote by $\real$ and $\n$ the sets of real numbers and nonnegative integers, and define $\n_{[a,b]}=\setb{a, a+1, \dots, b}$ where $a,b\in \n$ and $a<b$. Given a vector sequence $\{x_t\}$, define $\boldsymbol{x}_{[t_1,t_2]} = (x_{t_1}, x_{t_1+1}, \dots, x_{t_2}) = [x\tran_{t_1}, x\tran_{t_1+1}, \dots, x\tran_{t_2}]\tran$ , $t\in\n_{[t_1,t_2]}$. When $X_t$ are random vectors, $\boldsymbol{X}_{[t_1,t_2]}$ is a random process. We use $\bP$ to denote probability. A random vector $X \sim \mathcal{N}(\mu, \Sigma)$ follows a multivariate Gaussian distribution with mean $\mu$ and covariance $\Sigma$.
Given a vector $x\in \real^n$, $\|x\|$ denotes the euclidean norm and $\|x\|_P = \sqrt{x\tran P x}$, where $P\in\real^{n\times n}$ is a positive definite matrix.
The $n$ dimensional ball with radius $r$ and center $y$ is denoted by $\BB^n(r, y)=\setb{x\in \real^n : \|x-y\| \leq r}$. Denote the complement of set $A$ as $\setcomp{A}$ and $-B = \setb{-y: \forall y\in B}$. Given sets $A$ and $B$, define the Minkowski sum of $A$ and $B$ by $A\oplus B = \setb{x+y: x\in A,~ y\in B}$, and the Minkowski difference or Pontryargin difference of $A$ and $B$ by $A\ominus B=\setb{x:x+y\in A, \forall y\in B}$ \cite{kolmanovsky1998theory}. The Minkowski sum and difference satisfy the relation $(A\ominus B)\oplus B \subseteq A$.

\section{Problem Formulation}
In this section, we provide an overview of signal temporal logic and then formulate the STL verification problem for nonlinear stochastic systems.

\subsection{Signal Temporal Logic}
In this work, we use STL to specify the spatio-temporal properties of systems for safety verification. 
We consider the STL formula with a standard syntax
\begin{equation}
\label{eq: standard STL syntax}
    \varphi := \mathsf{T} {}\mid{} \pi {}\mid{} \neg\varphi {}\mid{} \varphi_1\wedge\varphi_2 {}\mid{} \varphi_1 \UU_{[t_1,t_2]} \varphi_2,
\end{equation}
where $\mathsf{T}$ denotes \textit{True}, $\pi:=(\mu(x)\geq 0)$ is a predicate, $\neg$ and $\wedge$ denote negation and conjunction, and $\UU_{[t_1,t_2]}$ is the temporal \textit{until} operator. An STL formula $\varphi$ is recursively constructed using the operators introduced above. We use $\pi \in \varphi$ to denote that the predicate $\pi$ is used to construct $\varphi$. Other operators can be constructed using these operators. For example, temporal \textit{eventually} operator $\lozenge_{[t_1,t_2]}\varphi = \mathsf{T} \UU_{[t_1,t_2]}\varphi$, temporal \textit{globally} operator $\square_{[t_1,t_2]}\varphi = \neg \lozenge_{[t_1,t_2]}\neg \varphi$, and disjunction $\varphi_1\vee \varphi_2 = \neg (\neg \varphi_1 \wedge \neg \varphi_2)$.

The boolean semantics of the STL formula are defined over the system trajectory $\boldsymbol{x}_{[t,\infty]}$ ~\cite{maler2004monitoring}.
Denote by $\boldsymbol{x}_{[t,\infty]} \models \varphi$ a trajectory $\boldsymbol{x}_{[t,\infty]}$ satisfies $\varphi$. The boolean semantics can be recursively defined by: 
$\boldsymbol{x}_{[t,\infty]} \models \pi \Leftrightarrow \mu(x_t) \geq 0$, 
$\boldsymbol{x}_{[t,\infty]} \models \neg \varphi \Leftrightarrow \neg(\boldsymbol{x}_{[t,\infty]} \models \varphi)$, 
$\boldsymbol{x}_{[t,\infty]} \models \varphi_1 \wedge \varphi_2 \Leftrightarrow \boldsymbol{x}_{[t,\infty]} \models \varphi_1 \wedge \boldsymbol{x}_{[t,\infty]} \models \varphi_2$,
$\boldsymbol{x}_{[t,\infty]} \models \varphi_1 \UU_{[t_1,t_2]}\varphi_2 \Leftrightarrow \exists \tau\in \n_{[t+t_1,t+t_2]}$, 
$\st \boldsymbol{x}_{[\tau,\infty]} \models \varphi_2 \wedge \forall \tau'\in \n_{[t,\tau]}, \boldsymbol{x}_{[\tau',\infty]} \models \varphi_1$. 

The horizon of an STL formula $\varphi$ is denoted as $h^\varphi$, which is the length of the trajectory that is required to determine the satisfaction of $\varphi$~\cite{belta2019formal}. $h^\varphi$ can be calculated recursively by: $h^\pi=0$, $h^{\neg \varphi} = h^\varphi$, $h^{\varphi_1 \wedge \varphi_2} = \max(h^\varphi_1, h^\varphi_2)$, $h^{\varphi_1 \UU_{[t_1,t_2]}\varphi_2} = t_2 + \max(h^\varphi_1, h^\varphi_2)$.

To simplify the analysis, we assume that all STL formulas are converted into a negation-free form, meaning the formulas do not contain any negations. This conversion is always possible by first transforming the formula into Negation Normal Form~\cite{fainekos2009robustness} and then introducing new predicates with reversed inequalities as needed~\cite[Proposition 2]{belta2019formal, sadraddini2015robust}.

\subsection{Stochastic Systems}
{\em Dynamics:}
Consider the discrete-time stochastic system 
\begin{equation}
\label{eq: stochastic dynamics}
     X_{t+1}=f(X_t,d_t,t)+w_t
\end{equation}
where $X_t\in \real^n$, $d_t\in \DD \subset \real^m$, $w_t\in \real^n$ are the state, input, and stochastic disturbance at time $t \in \n$. The input $d_t$ represents bounded disturbances, whose statistical property is unknown. We assume $f: \real^n\times\real^p\times \n \ra\real^n$ is globally Lipschitz (Assumption \ref{ass: Lipschitz f}). 
\begin{assumption}[Lipschitz]
\label{ass: Lipschitz f}
For $\forall t \in \n$, there exists a $L_t$, such that for all $x, y\in \real^n$, and $d\in\DD$, 
\begin{equation*}
    \|f(x,d,t) - f(y,d,t)\| \leq L_t\|x-y\|.
\end{equation*}
\end{assumption}

{\em Stochastic Disturbance:}
We assume the stochastic disturbance $w_t$ is sub-Gaussian, which covers a wide range of distributions in the real world including Gaussian distribution and uniform distribution with bounded support.

\begin{definition}
A random vector $ X \in \real^n$ is said to be sub-Gaussian if $\EE(X)=0$ and there exists a positive constant $ \sigma > 0 $ such that for any $\ell$ on the unit sphere, $\EE_X \left[ e^{\lambda \langle\ell, X\rangle} \right] \leq e^{\frac{\sigma^2 \lambda^2}{2}}$ holds for all $ \lambda \in \real $. Here, $ \sigma^2 $ is called variance proxy. We use $X\sim subG(\sigma^2)$ to denote $X$ is sub-Gaussian.
\end{definition}
\begin{assumption}\label{ass:subG}
    For system~\eqref{eq: stochastic dynamics}, 
    $w_t \sim subG(\sigma_t^2)$, where $\sigma_t>0$, $\forall t>0$.
\end{assumption}

\subsection{Problem Statement}
We consider the safety verification problem of the stochastic system \eqref{eq: stochastic dynamics} under the STL specification \eqref{eq: standard STL syntax}.
We first recall the STL satisfaction of a deterministic system under an STL specification.
\begin{definition}
Consider the deterministic version of the system dynamics
    \begin{equation}
    \label{eq: deterministic dynamics}
    x_{t+1}=f(x_t,d_t,t)
    \end{equation}
with a set of initial states $\XX_0$, and a bounded set of input $\DD$. Given an STL specification $\varphi$ with a bounded horizon $T$, system \eqref{eq: deterministic dynamics} is said to satisfy $\varphi$ if,
    \begin{equation}\label{eq: probability constraint}
        \forall x_0 \in \XX_0, \forall t\in\n_{[0,T]}, d_t \in \DD: \boldsymbol{x}_{[0,T]} \models \varphi.
    \end{equation}
\end{definition}

This definition can be conservative for the stochastic system \eqref{eq: stochastic dynamics} whose trajectories are often unbounded, inevitably leading to a violation of the STL specification. Therefore, we turn to a probabilistic version of STL satisfaction~\cite{sadigh2016safe, farahani2018shrinking}.
\begin{definition}
Consider the stochastic system \eqref{eq: stochastic dynamics} with a set of initial states $\XX_0$, and a bounded set of input $\DD$. Given an STL specification $\varphi$ with a bounded horizon $T$, and a probability tolerance $\delta \in [0,1]$, the system is said to satisfy $\varphi$ with $1-\delta$ guarantee, if
    \begin{equation}\label{eq: probability constraint}
        \forall x_0 \in \XX_0, \forall t\in\n_{[0,T]}, d_t \in \DD: \bP \big(\boldsymbol{X}_{[0,T]} \models \varphi \big) \geq 1-\delta.
    \end{equation}

\end{definition}

\begin{remark}
$\XX_0$ and $\DD$ may each contain only a single point. For instance, if $\XX_0 = \{x_0\}$ and $\DD = \{0\}$, this corresponds to considering a single initial state $x_0$ and no deterministic disturbance, which is a common setting in optimal control problems~\cite{vlahakis2024probabilistic}. 
\end{remark}

We are interested in the problem of verifying a stochastic system under an STL specification.
\begin{problem}[Verification]
\label{prob: verification}
Consider a stochastic system \eqref{eq: stochastic dynamics} under Assumptions \ref{ass: Lipschitz f}-\ref{ass:subG}, and an STL specification $\varphi$. Verify whether the system satisfies $\varphi$ with $1-\delta$ guarantee. 
\end{problem}

\section{Stochastic STL verification}\label{sec:main}
Our strategy to solve Problem \ref{prob: verification} is to convert the probabilistic STL satisfaction problem into a deterministic one with a tighter STL formula, as described in Section~\ref{sec: stl erosion} and Figure~\ref{fig: method}. Combining this strategy and a sharp bound on the deviation of a stochastic trajectory from its deterministic counterpart, we address Problem \ref{prob: verification} in Section~\ref{sec: probabilistic bound}.

\subsection{STL Erosion}
\label{sec: stl erosion}

\begin{figure}
\centering
\includegraphics[width =0.9\linewidth]{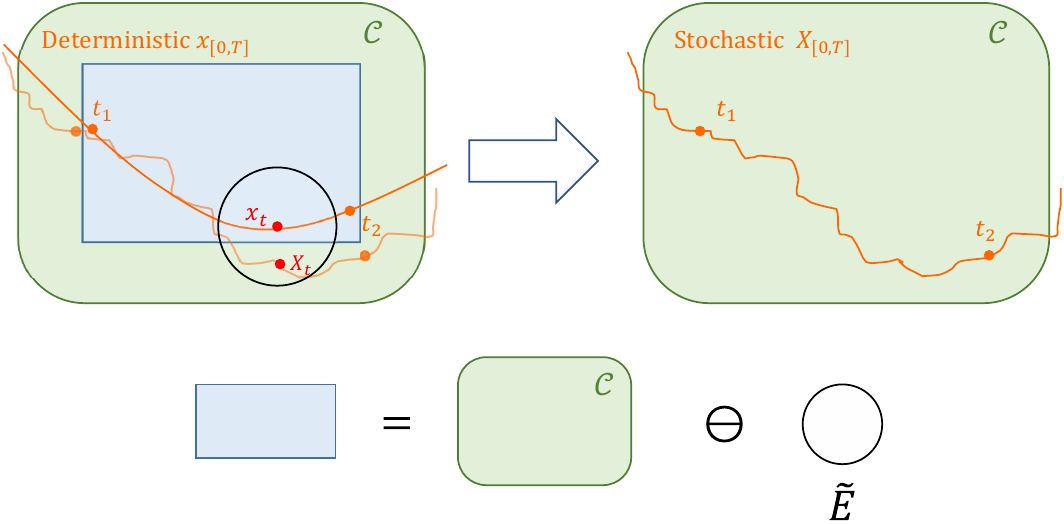}
\caption{An illustration of the STL erosion method. Here we consider a simple STL specification: $\varphi:=\square_{[t_1,t_2]}\pi$, where the super level set of the predicate $\pi$ is $\CC$. This STL specification requires the state to remain inside a safe set $\CC$ from $t_1$ to $t_2$. The predicate $\pi$ is eroded by $\tilde{E}$. If the deterministic trajectory can satisfy the tightened STL formula $\tilde \varphi = \square_{[t_1,t_2]}\tilde\pi$, then the stochastic trajectory would satisfy $\varphi$ with a high probability.}
\label{fig: method}
\end{figure}

The deterministic system~\eqref{eq: deterministic dynamics} can be viewed as a noise-free version of the stochastic system~\eqref{eq: stochastic dynamics}. We refer to a deterministic trajectory of \eqref{eq: deterministic dynamics} and a stochastic trajectory of \eqref{eq: stochastic dynamics} from the same initial state and subjected to the same sequence of $d_t$ as \textit{associated} trajectories. Our strategy relies on the intuition that the trajectory of \eqref{eq: stochastic dynamics} should fluctuate around and remain close to its associated trajectory of \eqref{eq: deterministic dynamics} with high probability.

Building on the set erosion strategy~\cite{liu2024safety}, we introduce the \textit{predicate erosion strategy}. 
For a predicate $\mu(\cdot)$, let $\CC$ be its superlevel set, \ie, $\CC=\setb{x\in \real^n:\mu(x)\geq 0}$. We shrink $\CC$ based on the amount of fluctuation to get a subset $\tilde{\CC}\subset\CC$. If for the deterministic state $x_t$, $x_t \in \tilde{\CC}$, then for the stochastic state $X_t$, $X_t\in \CC$ holds with a high probability since the fluctuation of the stochastic trajectory probably not exceed the margin $\CC \backslash \tilde{\CC}$. Therefore, we can verify whether the stochastic trajectory satisfies $\varphi$ with a high probability by verifying whether the deterministic trajectory satisfies $\varphi$ with the eroded predicates. 

To quantify the fluctuations of a stochastic trajectory around its deterministic counterpart, we formalize the concepts of stochastic fluctuation and stochastic deviation.

\begin{definition}
    Define $e_t = X_t - x_t$ as the stochastic fluctuation of the stochastic trajectory around the associated deterministic trajectory at time step $t$. Define $\|e_t\| = \| X_t-x_t\|$ as the stochastic deviation.
\end{definition}

Analogous to the reachability analysis of the system state, we introduce the probabilistic reachable set of the stochastic fluctuation. 

\begin{definition}[PRS] 
    Let $e_t$ be the stochastic fluctuation at time step $t$.
    A set $E_{\theta, t}$ is called a Probabilistic Reachable Set (PRS) of $e_t$ at probability level $1-\theta$, $\theta \in (0,1)$, if for any $x_0=X_0\in \XX_0$ and any $d_s \in \DD, 0\le s\le t$, it holds that
    \begin{equation}
        \bP(e_t\in E_{\theta, t})\geq 1-\theta.
    \end{equation}
\end{definition}

Note that the PRS of $e_t$ is not unique. If $E_{\theta, t}$ is a PRS, then $E'_{\theta, t}$ is also a PRS if $E_{\theta, t}\subseteq E'_{\theta, t}$ We say $E_{\theta, t}$ is tighter than $E'_{\theta, t}$ if $E'_{\theta, t} \subseteq E_{\theta, t}$. We present a tight PRS in \ref{sec: probabilistic bound}.

For ease of analysis, we introduce the following sets:
\begin{equation} \label{eq: E_theta}
    \tilde{E}_\theta = \bigcup_{t=0}^{T} E_{\theta, t},
\end{equation}
\begin{equation} \label{eq: E_theta T}
    \boldsymbol{\tilde E}_{\theta} = \underbrace{\tilde{E}_\theta \times \tilde{E}_\theta \times \dots \times \tilde{E}_\theta}_{T \text{ times}}.
\end{equation}
At any time step \( t \), the stochastic fluctuation \( e_t \) belongs to the set \( \tilde{E}_\theta \) with probability at least \( 1 - \theta \). By union bound, the probability that \( e_t \) belongs to \( \tilde{E}_\theta \) for all time steps is
\begin{equation}
   \mathbb{P}(e_t \in \tilde{E}_\theta, \forall t \in \mathbb{N}_{[0,T]}) = \mathbb{P}(\boldsymbol{e}_{[0,T]} \in \boldsymbol{\tilde E}_{\theta}) \geq 1 - T\theta
\end{equation}

Using the PRS, we can tighten the predicates to eliminate the probabilistic term in \eqref{eq: probability constraint}.

\begin{proposition}[Predicate erosion]
    \label{prop: predicate erosion}
    Assume $E_{\theta, t}$ is a PRS of $e_t$. Let $\tilde{E}_\theta$ be as defined in \eqref{eq: E_theta}. Consider a predicate $\pi = (\mu(\cdot)\geq 0)$ with superlevel set $\CC$. Let $\tilde \pi$ be a tighter version of $\pi$, \st $x_t \models \tilde \pi \Leftrightarrow x_t \in \CC \ominus \tilde{E}_\theta$, then for any time step $t$, $t \in \n_{[0, T]}$, $x_t \models \tilde \pi \implies x_t + e_t \models \pi, \forall e_t \in \tilde{E}_\theta$.
\end{proposition}
\begin{proof}
    $x_t\models \tilde \pi \implies x_t \in \CC \ominus \tilde E_{\theta} \implies \forall e_t \in \tilde E_{\theta}, x_t + e_t \in (\CC \ominus \tilde E_{\theta, t})\oplus \tilde E_{\theta} \subseteq \CC$. Therefore $ x_t + e_t \models \pi, \forall e_t\in \tilde E_{\theta}$.
\end{proof}
Proposition~\ref{prop: predicate erosion} extends \cite{vlahakis2024probabilistic} which considers predicate tightening for affine predicates. 
In Proposition~\ref{prop: predicate erosion}, we erode $\CC$ with $\tilde{E}_\theta$, as verifying an STL formula may require evaluating a predicate at any timestep along a trajectory. Compared to the Set Erosion method for the safe set specification~\cite{liu2024safety}, the STL specification involves both spatial and temporal constraints, requiring us to shrink $\CC$ for all time steps using the union of the PRSs, as opposed to the time-varying safe set in~\cite{liu2024safety}. The following theorem shows that the stochastic STL satisfaction problem can be converted into a deterministic one by eroding every predicate in an STL formula.

\begin{theorem}[STL formula erosion]
\label{thm: STL erosion}
    Consider associated trajectory $\boldsymbol{X}_{[0,T]}$ and $\boldsymbol{x}_{[0,T]}$. Let $\tilde{E}_\theta$ be as defined in \eqref{eq: E_theta}, and $\boldsymbol{\tilde E}_{\theta}$ be as defined in \eqref{eq: E_theta T}. Given an STL formula $\varphi$ constructed with predicates $\pi_1, \pi_2, \dots, \pi_m$, and corresponding superlevel sets $\CC_1, \CC_2, \dots, \CC_m$. For every predicate $\pi_i,i \in \n_{[1,m]}$, we substitute $\pi_i$ with $\tilde \pi_i$, where the superlevel set of $\tilde \pi_i$ is $\CC_i \ominus \tilde{E}_\theta$ and keep all other operators unchanged to get $\tilde{\varphi}$, a eroded version of $\varphi$. Then $\boldsymbol{x}_{[0,T]} \models \tilde{\varphi} \implies \bP(\boldsymbol{X}_{[0,T]}\models \varphi) \geq 1-T\theta$
\end{theorem}
\begin{proof}
    By definition, it suffices to show $\boldsymbol{x}_{[0,T]} \models \tilde{\varphi} \implies \boldsymbol{x}_{[0,T]} +  \boldsymbol{e}_{[0,T]}\models \varphi, \forall \boldsymbol{e}_{[0,T]} \in \boldsymbol{\tilde E}_{\theta}$. Since the STL formulas are recursively defined, we prove by induction.

    \textit{Base Case}: By Proposition~\ref{prop: predicate erosion}, $\forall t \in \n_{[0,T]}$, $\boldsymbol{x}_{[t,T]} \models \tilde \pi_i\implies \boldsymbol{x}_{[t,T]} +  \boldsymbol{e}_{[t,T]} \models \pi_i, \forall \boldsymbol{e}_{[t,T]} \in \boldsymbol{\tilde E}_{\theta [t, T]}$. 

   \textit{Induction Hypothesis}: For any STL sub-formulas $\varphi_j$ and their eroded version $\tilde \varphi_j$, assume $\forall t \in \n_{[0,T]}$, $\boldsymbol{x}_{[t,T]} \models \tilde \varphi_j\implies \boldsymbol{x}_{[t,T]} +  \boldsymbol{e}_{[t,T]} \models \varphi_j, \forall \boldsymbol{e}_{[t,T]} \in \boldsymbol{\tilde E}_{\theta [t, T]}.$ 

   \textit{Induction Step}: We discuss logical operators and temporal operators respectively:

    \begin{enumerate}
        \item \textit{Logical operators:}  Let $\varphi' = \varphi_1 \wedge \varphi_2$ and assume $\boldsymbol{x}_{[t,T]} \models \tilde{\varphi}'$, which is equivalent to $\boldsymbol{x}_{[t,T]}\models\tilde \varphi_1 \wedge \boldsymbol{x}_{[t,T]}\models \tilde\varphi_2$. This implies $\boldsymbol{x}_{[t,T]} +  \boldsymbol{e}_{[t,T]} \models \varphi_1 \wedge \boldsymbol{x}_{[t,T]} +  \boldsymbol{e}_{[t,T]} \models \varphi_2$, which is equivalent to $\boldsymbol{x}_{[t,T]} +  \boldsymbol{e}_{[t,T]} \models \varphi'$, $\forall \boldsymbol{e}_{[t,T]} \in \boldsymbol{\tilde E}_{\theta [t, T]}$. We can similarly prove the induction step for the disjunctive operator. 
        \item \textit{Temporal Operators:} Let $\varphi' = \varphi_1\UU_{[t_1,t_2]}\varphi_2$ and assume $\boldsymbol{x}_{[t,T]} \models \tilde{\varphi}'$, which is equivalent to $\exists \tau \in \n_{[t+t_1,t+t_2]}, \boldsymbol{x}_{[\tau,T]} \models\tilde{\varphi}'_2 \wedge \forall \tau' \in \n_{[t, \tau]}, \boldsymbol{x}_{[\tau',T]} \models \tilde{\varphi}'_1$. This implies that, for the same $\tau$, $\boldsymbol{x}_{[\tau,T]} + \boldsymbol{e}_{[\tau,T]} \models {\varphi}'_2 \wedge \forall \tau' \in \n_{[t, \tau]}, \boldsymbol{x}_{[\tau',T]} + \boldsymbol{e}_{[\tau',T]} \models {\varphi}'_1$, which is equivalent to $\boldsymbol{x}_{[t,T]} + \boldsymbol{e}_{[t,T]} \models \varphi', \forall \boldsymbol{e}_{[t,T]} \in \boldsymbol{\tilde E}_{\theta [t, T]}$. We can similarly prove the induction step for other temporal operators.
        
    \end{enumerate}
    
    By induction, if $\boldsymbol{x}_{[0,T]} \models \tilde{\varphi}$, then $\boldsymbol{x}_{[0,T]} +  \boldsymbol{e}_{[0,T]}\models \varphi, \forall \boldsymbol{e}_{[0,T]} \in \boldsymbol{E}_{\theta,T}$.
\end{proof}

An illustration of the proposed method is shown in Figure~\ref{fig: method}. According to Theorem~\ref{thm: STL erosion}, instead of verifying whether a stochastic trajectory satisfies an STL formula with a certain probability, one can verify whether a corresponding deterministic trajectory satisfies a tighter STL formula. Consequently, the stochastic STL verification problem (Problem~\ref{prob: verification}) reduces to a deterministic verification problem, which can be solved by any existing methods.

\begin{theorem}[STL verification of stochastic systems]\label{thm: verification}
    Consider the stochastic system \eqref{eq: stochastic dynamics} and the associated deterministic system \eqref{eq: deterministic dynamics} with initial set $\XX_0$ and bounded input set $\DD$. Given an STL specification $\varphi$ with bounded horizon $T$, $\delta \in (0,1)$, define $\tilde{E}_\theta$ as in \eqref{eq: E_theta} with $\theta = \delta/T$ and construct $\tilde \varphi$ as in Theorem~\ref{thm: STL erosion}. If \eqref{eq: deterministic dynamics} satisfies $\tilde{\varphi}$, then the stochastic system \eqref{eq: stochastic dynamics} satisfies $\varphi$ with $1-\delta$ guarantee.
\end{theorem}

\begin{proof}
    The deterministic system satisfies $\tilde{\varphi}$ implies that, for every initial state $x_0 \in \XX_0$ and every bounded disturbance $d_t \in \DD$, it holds that $\boldsymbol{x}_{[0,T]} \models \tilde \varphi$. By Theorem~\ref{thm: STL erosion}, for every initial state $x_0 \in \XX_0$ and every bounded disturbance $d_t \in \DD$, $\bP(\boldsymbol{X}_{[0,T]}\models \varphi) \geq 1-T\theta = 1-\delta$. Therefore the stochastic system satisfies $\varphi$ with $1-\delta$ guarantee.
\end{proof}

\subsection{STL verification}
\label{sec: probabilistic bound}
To effectively apply the STL erosion strategy, a tight PRS of the stochastic fluctuation is the key. Next, we utilize recent results on bounding stochastic derivation to obtain a tight PRS~\cite{liu2024probabilistic,jafarpour2024probabilistic}. 
\begin{proposition}\label{prop: stochastic deviation}\cite{liu2024probabilistic}
Let $\boldsymbol{X}_{[0,\infty]}$ be the trajectory of the stochastic system \eqref{eq: stochastic dynamics} and $\boldsymbol{x}_{[0,\infty]}$ be the trajectory of the deterministic system \eqref{eq: deterministic dynamics} with the same initial state $x_0 \in \XX_0$ and the same input sequence $\boldsymbol{d}_{[0, \infty]}$. Then for any $\varepsilon \in (0,1)$, $\theta \in (0,1)$, and $t\geq 0$,
\begin{equation}\label{eq: bound r}
 \bP\Big( \|X_t-x_t\|\leq r_{\theta, t} \Big) \geq 1-\theta,
\end{equation}
 where 
$r_{\theta, t} = \sqrt{\Psi_t(\varepsilon_1n+\varepsilon_2\log(1/\theta))}$, $\psi_t=\prod_{k=0}^{t}L_k^{2}$,\quad
$\Psi_t=\psi_{t-1}\sum_{k=0}^{t-1}\sigma_{k}^2\psi_k^{-1}$, 
$\varepsilon_1=\frac{2\log(1+2/\varepsilon)}{(1-\varepsilon)^2}, \varepsilon_2=\frac{2}{(1-\varepsilon)^2}$.
\end{proposition}

The bound in \eqref{eq: bound r} holds for both the Euclidean norm and the weighted norm \( \|\cdot \|_P \)~\cite[Section V-D]{jafarpour2024probabilistic}.
The above bound scales logarithmically with $T$ and $1/\delta$, which is tight for stochastic systems \cite{liu2024probabilistic}. By Proposition~\ref{prop: stochastic deviation}, the ball $\BB(r_{\theta, t}, 0)$ (or an ellipsoid when using the weighted norm) serves as a tight PRS of $e_t$.

With the tight PRS of $e_t$, we can combine it with any existing methods for STL verification for deterministic systems, and leverage the STL erosion strategy stated in Theorem~\ref{thm: STL erosion} to verify STL satisfaction of the stochastic system.

\begin{theorem}\label{thm: overall}
        Consider the stochastic system \eqref{eq: stochastic dynamics} and the associated deterministic system \eqref{eq: deterministic dynamics} with initial set $\XX_0$, and the bounded input set $\DD$. Given an STL specification $\varphi$ with bounded horizon $T$, $\delta \in (0,1)$, let $\BB(r_{\delta, t}, 0)$ serve as the PRS of $e_t$ in Theorem~\ref{thm: verification}, where $r_{\delta, t} = \sqrt{\Psi_t(\varepsilon_1n+\varepsilon_2\log(T/\delta))}$. If the deterministic system satisfies $\tilde{\varphi}$, then the stochastic system satisfies $\varphi$ with $1-\delta$ guarantee.
\end{theorem}

\begin{proof}
    Plug the bound stated in Proposition~\ref{prop: stochastic deviation} into Theorem~\ref{thm: verification}, then Theorem~\ref{thm: overall} follows.
\end{proof}

Worst-case analysis is used to verify the safety of systems under bounded disturbances~\cite{prajna2007framework}. It can also be applied to systems with unbounded noise for probabilistic verification by treating the noise as bounded noise with a high probability. This idea is used in \cite{vlahakis2024probabilistic} to handle probabilistic STL constraints for linear stochastic systems with affine predicates. The same idea can be readily extended to nonlinear stochastic systems.
By~\cite[Section IV-B]{liu2024safety},
\begin{equation}\label{eq: sd by worst}
    \|X_t-x_t\|\leq \sqrt{\psi_{t-1}}\sum_{k=0}^{t-1}\sigma_{k}\sqrt{\psi_k^{-1}(\varepsilon_1n+\varepsilon_2\log\frac{T}{\delta})}
\end{equation}
holds for all $t\in\n_{[0,T]}$ with probability at least $1-\delta$. Denote the right-hand side as $r_{\delta,t}^{\text w}$. It is shown that $r_{\delta,t}^{\text w}$ is always greater than $r_{\delta,t}$ \cite{liu2024safety}. Thus, using the worst-case bound $r_{\delta,t}^{\text w}$ in STL erosion is inherently more conservative than Theorem~\ref{thm: overall}.

\section{Numerical examples}
\label{sec: experiments}
We illustrate the proposed method through two numerical examples, one on linear dynamics and one on nonlinear dynamics. 

\subsection{Double Integrator}
\label{sec: double integrator}

Consider the discrete-time double integrator system: 
\begin{equation}\label{eq: double integrator}
    X_{t+1} = 
    \underbrace{
    \begin{bmatrix}
    1 & 0 & \Delta t & 0 \\
    0 & 1 & 0 & \Delta t \\
    0 & 0 & 1 & 0 \\
    0 & 0 & 0 & 1
    \end{bmatrix}}_{\textbf{A}}
    X_t +
    \underbrace{
    \begin{bmatrix}
    0 & 0 \\
    0 & 0 \\
    \frac{\Delta t}{m} & 0 \\
    0 & \frac{\Delta t}{m}
    \end{bmatrix}}_{\textbf{B}}
    (u_t + d_t) + w_t,
\end{equation}
where $X_t \in \mathbb{R}^4$ represents the state vector $[p_x, p_y, v_x, v_y]\tran$, $\Delta t$ is the discretization step size, $u_t$ is the control input, $d_t\in\DD$ is the bounded disturbance, $w_t \sim \mathcal{N}(0, \Sigma)$ is the noise term. We set $\Delta t = 0.01$, $\Sigma = \Delta t\cdot 0.04 \cdot I_4$, and mass $m=0.1$. The set of initial states is $\XX_0 = [0.3, 0.5]\times[0.3,0.5]\times\{0\}\times\{0\}$.

\begin{figure}
    \centering
    \includegraphics[width =0.49\linewidth]{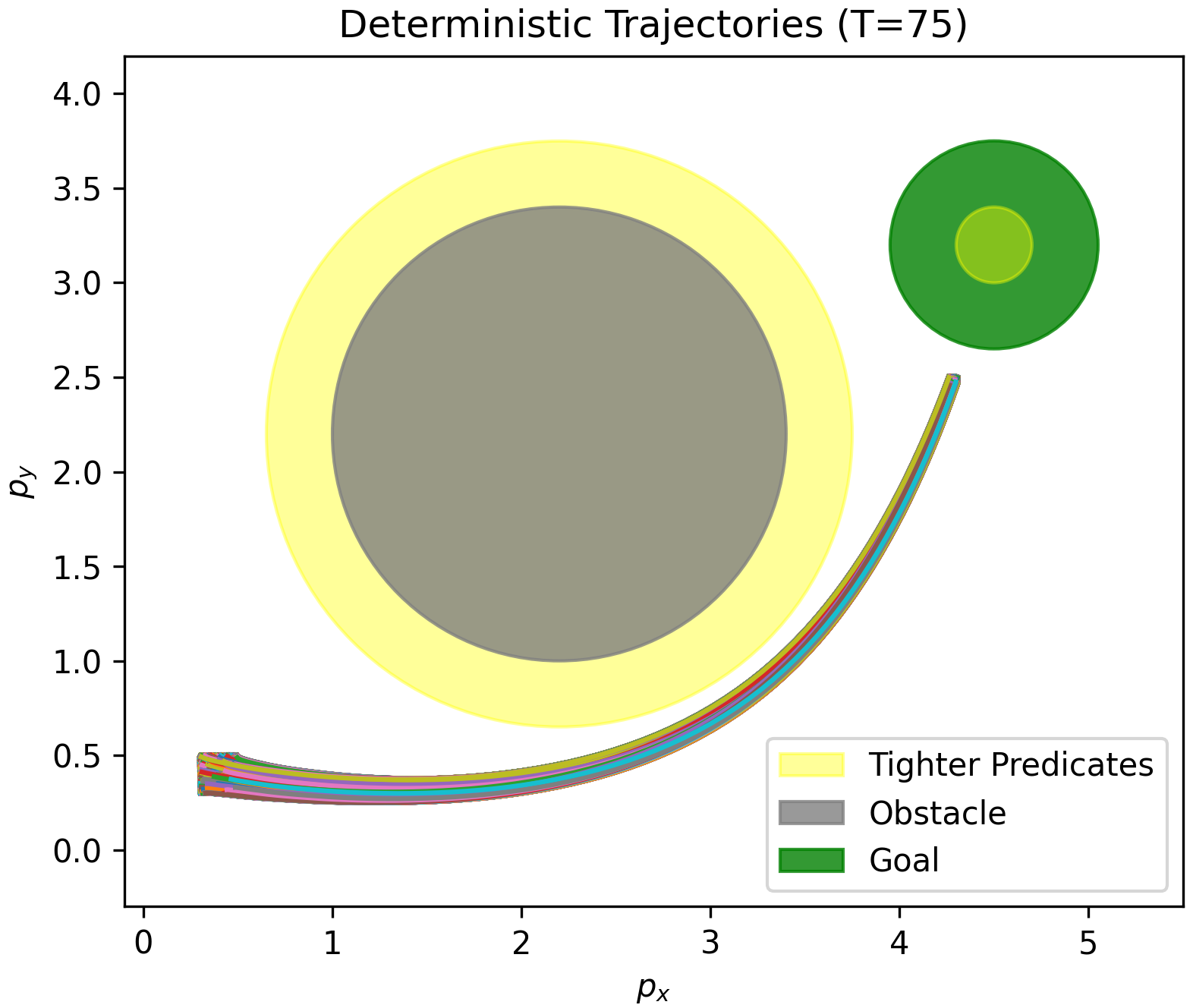}
    \includegraphics[width =0.49\linewidth]{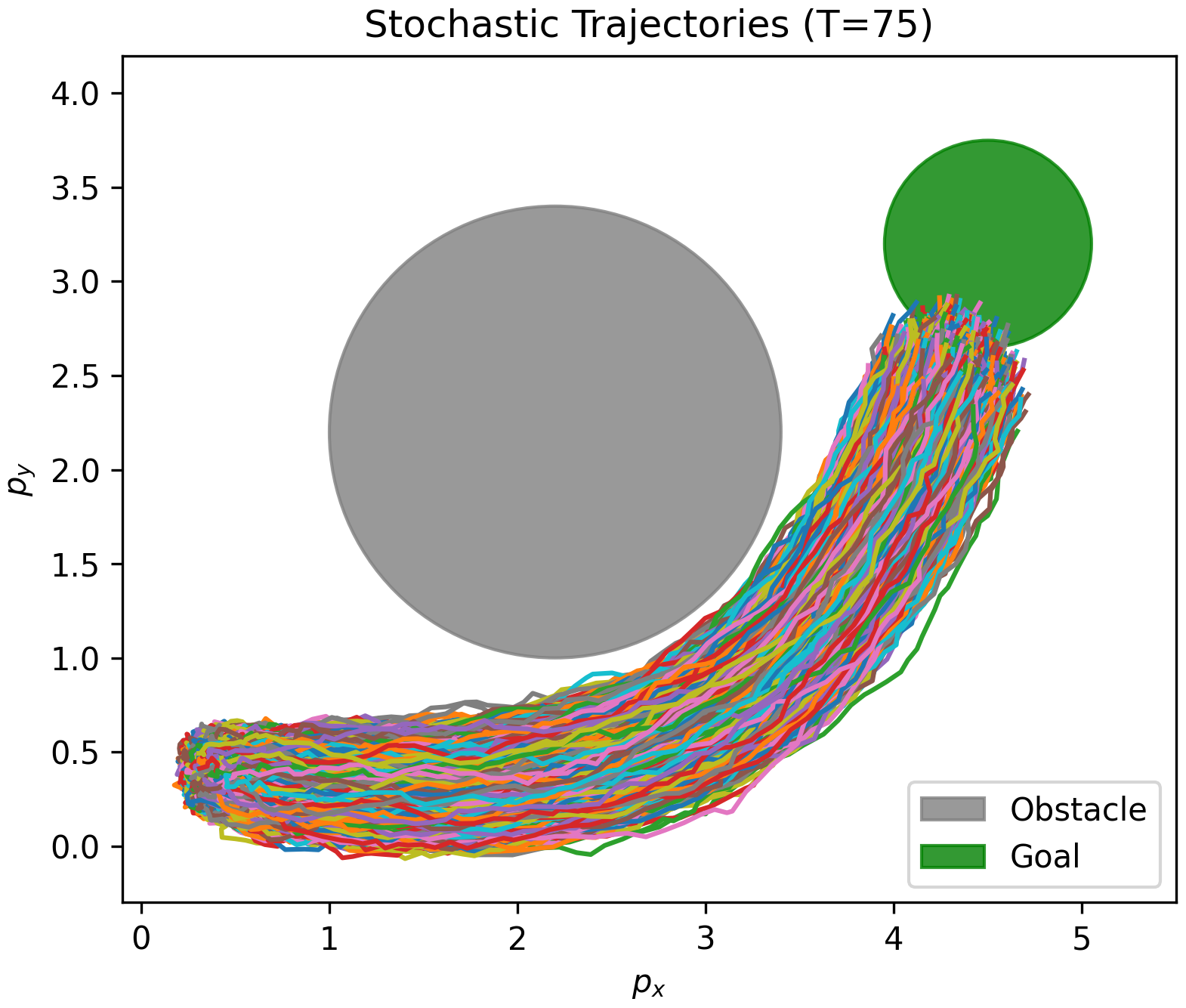}\\
    \vspace{0.2cm}
    \includegraphics[width =0.49\linewidth]{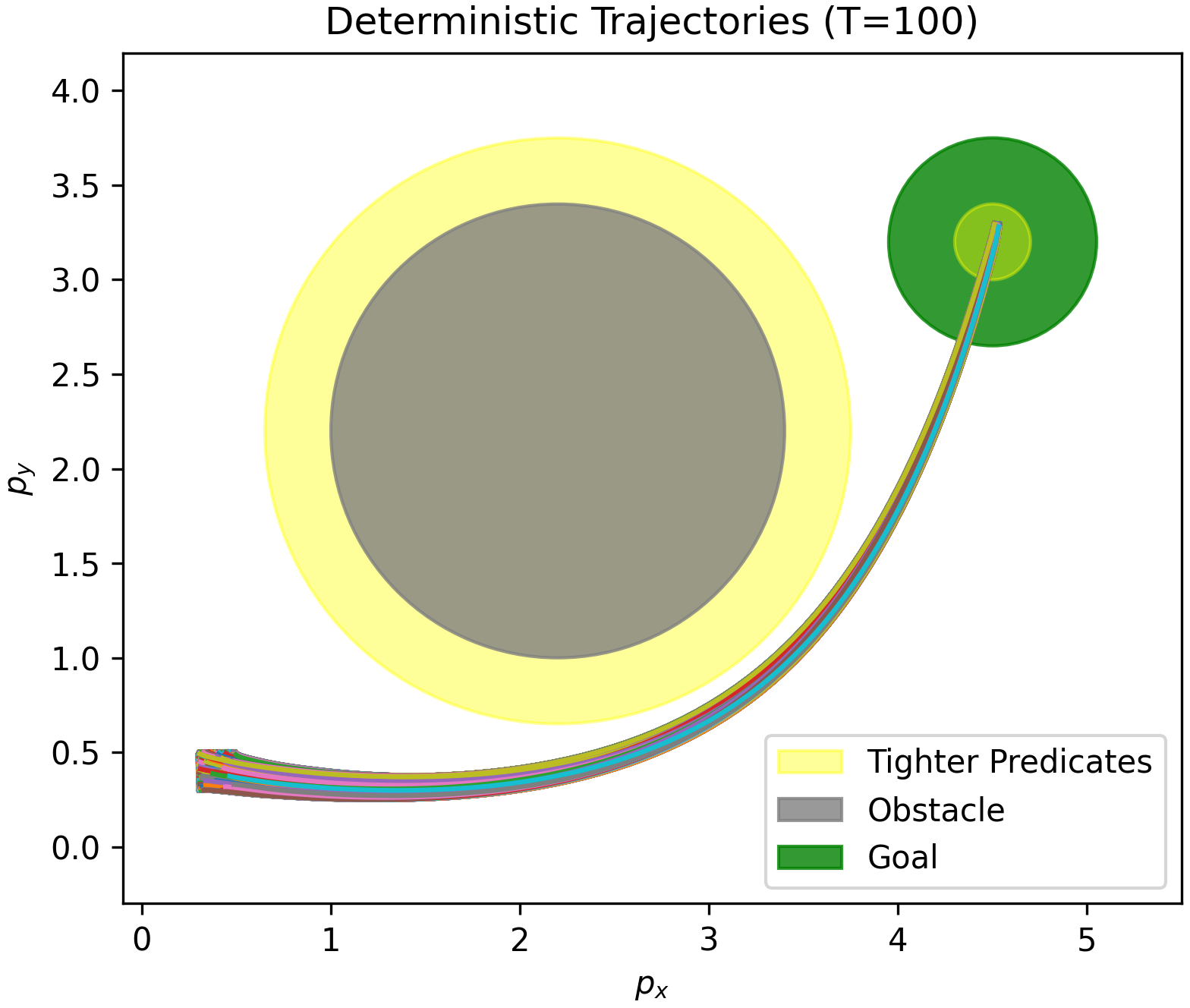}
    \includegraphics[width =0.49\linewidth]{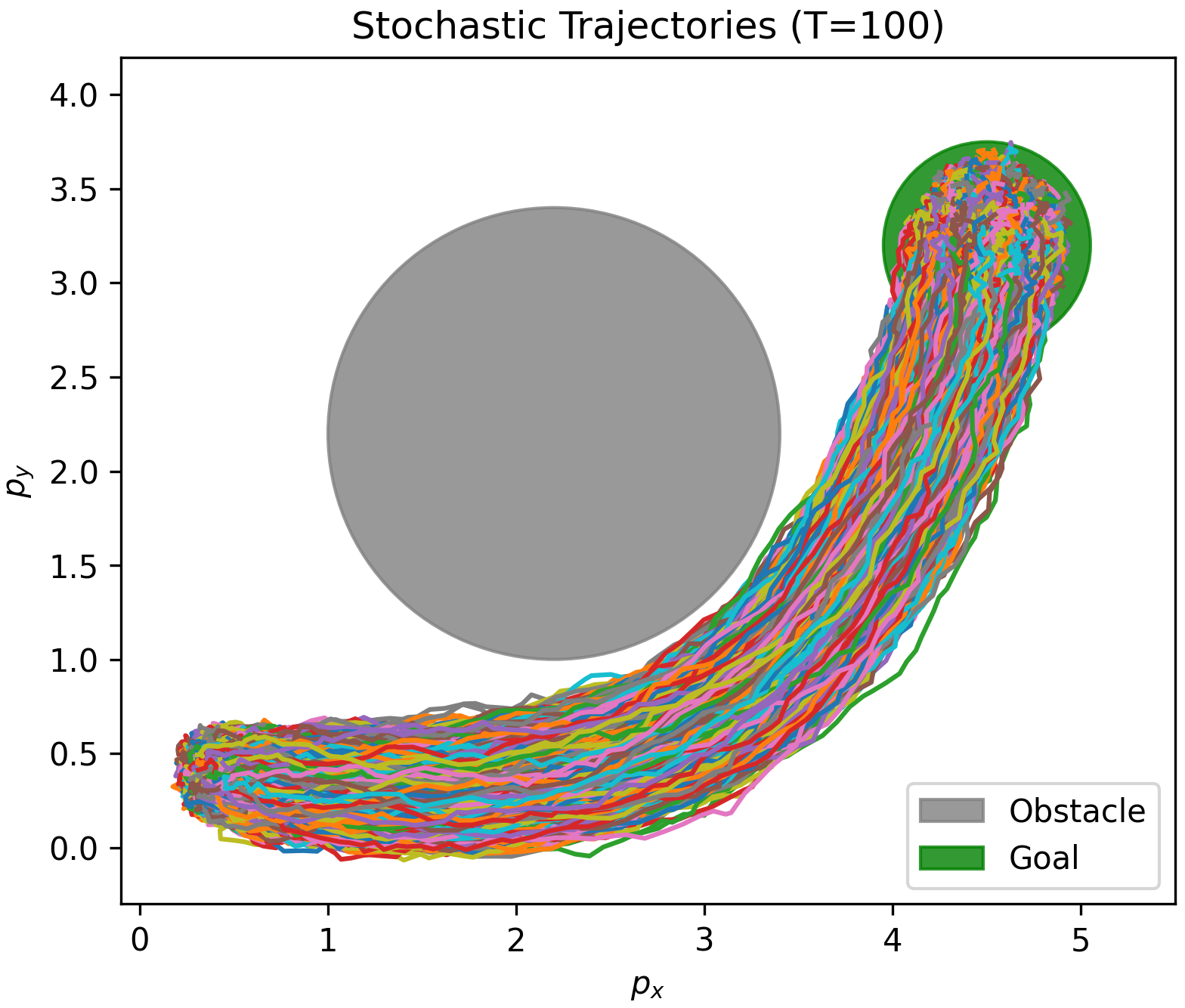}
    \caption{Stochastic STL verification of the double integrator system~\eqref{eq: double integrator} with $1-10^{-4}$ guarantee. \textbf{Left:} Stochastic STL verification using our STL erosion strategy. The gray circle and the green circle represent the obstacle and the goal area respectively. The corresponding eroded predicates are represented as yellow areas. Each curve is an independent trajectory of the deterministic system. \textbf{Right:} Each curve is an independent trajectory of the stochastic system.}
    \label{fig: double integrator}
\end{figure}

The task is to reach a goal area and stay in it for some time while avoiding obstacles, as shown in Figure~\ref{fig: double integrator}. The task horizon is $T=100$. To specify this task using an STL formula, we define two predicates, $\pi_{\text{obs}}:= \big(\mu_{\text{obs}}(x) = (p_x-2.2)^2+(p_y-2.2)^2 - 1.2^2 \geq 0\big)$, 
and $\pi_{\text{goal}}:= \big(\mu_{\text{goal}}(x) = (p_x-4.9)^2+(p_y-3.2)^2 - 0.55^2 \leq 0\big)$. 
The STL specification is $\varphi = (\square_{[0,T]}\pi_{\text{obs}})\wedge(\lozenge_{[0,T-10]}\square_{[0,10]}\pi_{\text{goal}})$, which requires the position to be inside the goal area continuously for 10 steps, and to be always collision-free with the obstacles.

We train a neural network to generate a reference trajectory for the deterministic system to complete the task following \cite{meng2023signal} and then use a static feedback gain $K = \begin{bmatrix}
    -2&0&-1&0\\
    0&-2&0&-1
\end{bmatrix}$ to track the reference trajectory. To get a small Lipschitz constant, we use weighted norm in~\eqref{ass: Lipschitz f}. We construct a semi-definite program to search for the best weight $P$ \cite{fan2017simulation}. The bound on stochastic fluctuation is given by the weighted norm $\|\cdot\|_P$, representing an ellipsoid in 4-dimensional space. Since all the predicates are specified in the $p_x - p_y$ plane, we project the ellipsoid to this plane to do predicate erosion.

Our goal is to verify whether the stochastic system with initial set $\XX_0$ and bounded disturbance $\DD = [-0.1, 0.1]\times [-0.1, 0.1]$ would satisfy $\varphi$ with probability $1-10^{-4}$. We calculate $\tilde{E}$ using our bound $r = r_{\delta, t} = 0.644$ (Theorem~\ref{thm: overall}). The result is visualized in Figure~\ref{fig: double integrator}. The deterministic system is verified to satisfy $\tilde \varphi$ by the algorithms implemented in CORA~\cite{roehm2016stl, lercher2024using, Althoff2015ARCH}. By Theorem~\ref{thm: overall}, the stochastic system satisfies $\varphi$ with probability at least $1-10^{-4}$. We simulate $10^5$ trajectories for both the deterministic system and the stochastic system to validate our method. All the sampled deterministic trajectories do not intersect with the obstacle enlarged by $\tilde{E}$, and stay inside the shrunk goal area for over 10 steps. Meanwhile, all the sampled stochastic trajectories satisfy the original STL formula $\varphi$, validating our strategy.

\subsection{Nonlinear Unicycle}
Consider the following nonlinear kinematic unicycle
\begin{equation}
    \begin{split}
        X_{t+1} &= X_t + \Delta t
        \begin{bmatrix}
            v_t\cos(\theta_t) \\
            v_t\sin(\theta_t) \\
            \omega_t + d_t
        \end{bmatrix}
        + w_t, 
    \end{split}
\end{equation}
where $X_t \in \mathbb{R}^3$ represents the state vector $[p_x, p_y,\theta]\tran$, $d_t\in \DD$ is the bounded disturbance, $w_t$ is the stochastic noise, and $v_t$ and $\omega_t$ are control inputs. $\Delta t$ is the discretization step size set to $0.05$. 

\begin{figure}
    \centering
    \includegraphics[width = 0.49\linewidth]{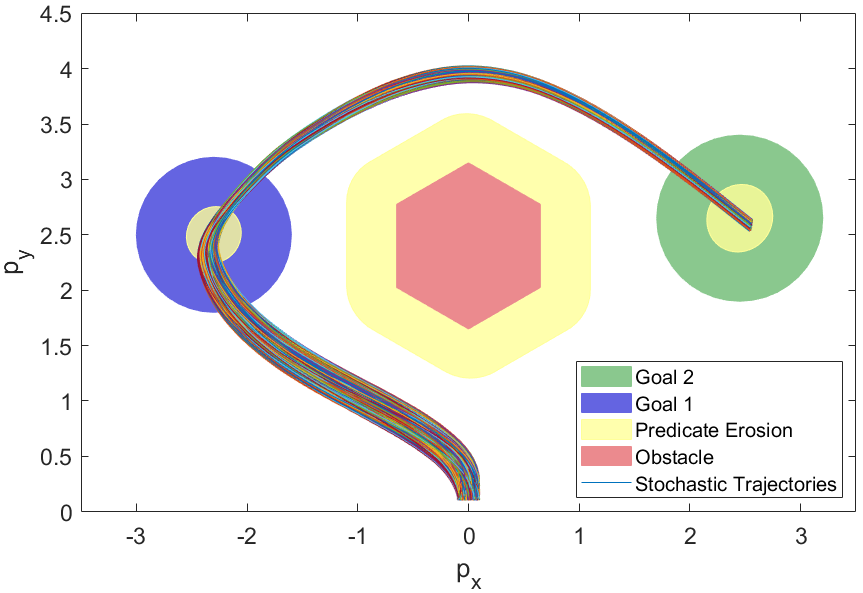}
    \includegraphics[width = 0.49\linewidth]{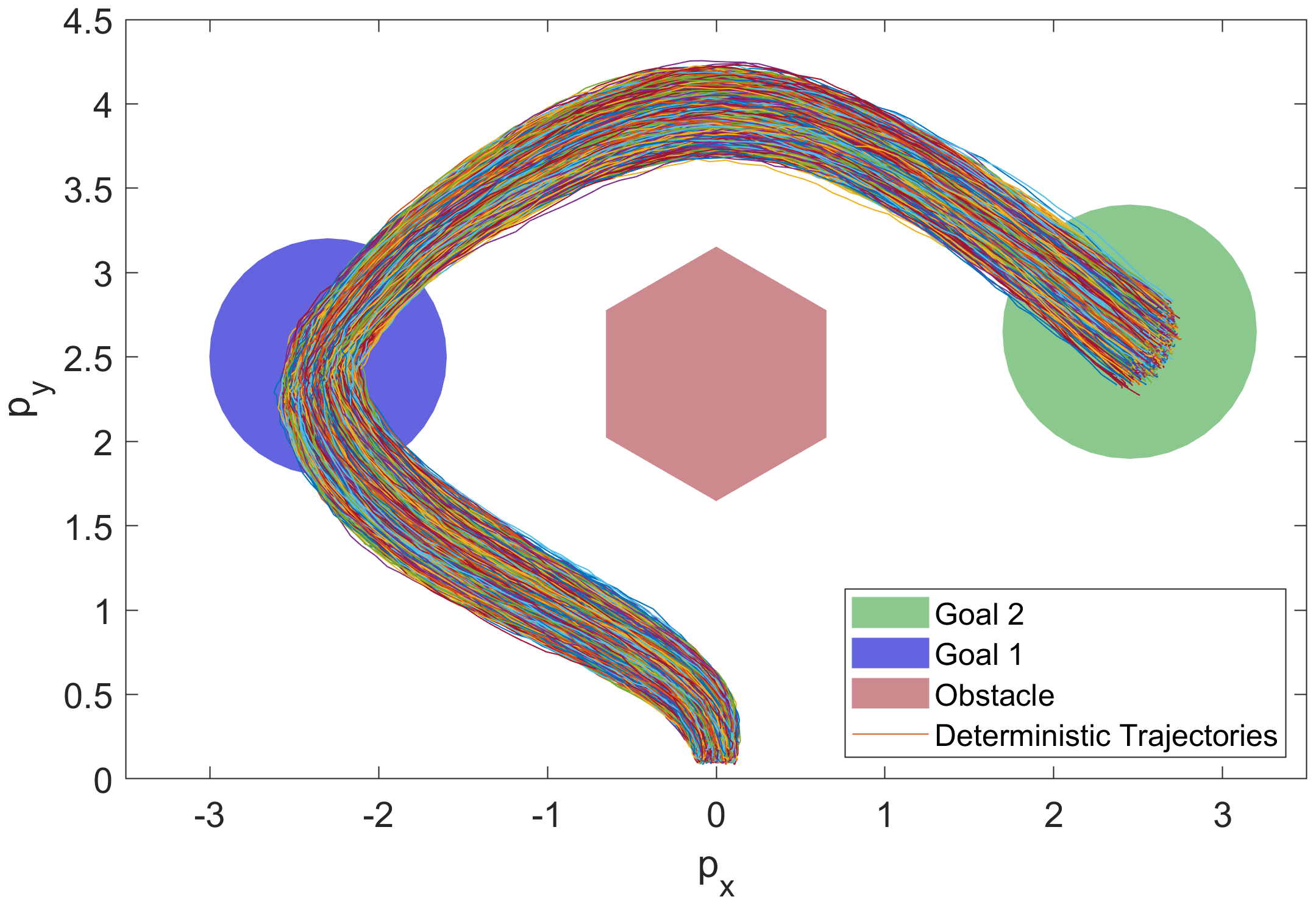}
    \caption{Stochastic STL verification of the unicycle system with $1-10^{-4}$ guarantee. \textbf{Left:} Stochastic STL verification using our STL erosion strategy. The blue circle represents the first goal area. The green circle represents the second goal area. The red hexagon is the obstacle. The corresponding eroded predicates are represented as yellow areas. Each curve is an independent trajectory of the deterministic system. \textbf{Right:} Each curve is an independent trajectory of the stochastic system.}
    \label{fig: unicycle}
\end{figure}

The task is to pass the first goal area (Goal 1) and finally enter the second goal area (Goal 2) while avoiding an obstacle in the middle, as shown in Figure~\ref{fig: unicycle}. Define three predicates, $\pi_{\text{goal}_1} = \big(\mu_{\text{goal}_1}(x) = (p_x+2.3)^2+(p_y-2.5)^2 - 0.7^2 \leq 0\big)$, $\pi_{\text{goal}_2} = \big(\mu_{\text{goal}_2}(x) = (p_x-2.45)^2+(p_y-2.65)^2 - 0.75^2 \leq 0\big)$, and $\pi_{\text{obs}}$ is the complement of the inscribed hexagon of the circle $(p_x-0)^2+(p_y-2.4)^2 - 1.2^2 \geq 0$. We can specify the task using the following STL formula, $\varphi = (\square_{[0,T]}\pi_{\text{obs}})\wedge(\lozenge_{[0,T]}\pi_{\text{goal}_2})\wedge(\neg \pi_{\text{goal}_2}\UU_{[0,T]} \pi_{\text{goal}_1})$.

Similar to Section~\ref{sec: double integrator}, we generate a reference trajectory $t\mapsto(p^*_x, p^*_y, \theta^*, v^*,\omega^*)$ for this task and use the a feedback tracking controller to track the reference trajectory:
$
    v_t = v^* + K_x\big(\cos\theta (p^*_x - p_x) + \sin\theta (p^*_y - p_y)\big),
    \omega_t = \omega^* + K_y\big(-\sin\theta (p^*_x - p_x) + \cos\theta (p^*_y - p_y) \big)
                + K_\theta(\theta^*-\theta).
$
The Lipschitz constant is estimated by sampling.

Our goal is to verify whether the stochastic system with initial set $\XX_0 = [-0.1, 0.1]\times[0.1, 0.3]\times[\pi/2 - 0.1, \pi/2+0.1]$, bounded disturbance $\DD = [-0.02, 0.02]$ and stochastic disturbance $w_t \sim \mathcal{N}(0, \Sigma)$, $\Sigma =\Delta t \cdot 0.001 \cdot I_3$ would satisfy $\varphi$ with probability $1-10^{-4}$. 

We first use our bound (Theorem~\ref{thm: overall}) to erode the STL formula. The result is visualized in Figure~\ref{fig: unicycle}. The deterministic system is verified to satisfy the tighter STL formula using CORA~\cite{Althoff2015ARCH}. We sample $10^5$ trajectories for both the deterministic system and the stochastic system, all of which satisfy the tighter STL formula and the original STL formula, respectively. 

Finally, we compare our method with the worst-case analysis described at the end of Section~\ref{sec:main}. The latter is significantly more conservative. In this example, $r_{\delta, t}^{\text w} \approx 6.91$, while our method yields $r_{\delta, t}=0.63$. As a result, the obstacle covers all goal areas and the goal areas become empty sets. The verification algorithm returns false due to conservativeness.

\section{Concluding remarks}
We propose an STL erosion strategy for probabilistic STL verification of discrete-time nonlinear stochastic systems under sub-Gaussian disturbances. Leveraging a tight bound on stochastic deviation, our strategy reduces the probabilistic verification problem into a deterministic verification problem with a tightened STL specification which can then be solved using existing deterministic methods. Our strategy is validated on both linear and nonlinear systems. Control synthesis for nonlinear stochastic systems with probabilistic STL specifications based on our results is an interesting and promising future direction.

\bibliographystyle{IEEEtran}        
\bibliography{references}

\end{document}